# Broadband Femtosecond Transient Absorption Spectroscopy for CVD MoS$_2$ monolayer

Shrouq H. Aleithan[1], Maksim Y. Livshits[2,3], Sudiksha Khadka[1], Jeffrey J. Rack[2,3], Martin Kordesch[1] and Eric Stinaff[1,*]

[1] *Department of Physics and Astronomy, Ohio University, Athens, Ohio, OH 45701, USA*

[2] *Department of Chemistry, Ohio University, Athens, Ohio, OH 45701, USA*

[3] *Department of Chemistry and Chemical Biology, University of New Mexico, Albuquerque, NM 87131, USA*

*Carrier dynamics in monolayer MoS$_2$ have been investigated using broadband femtosecond transient absorption spectroscopy (FTAS). A tunable pump pulse was used while a broadband probe pulse revealed ground and excited state carrier dynamics. Interestingly, for pump wavelengths both resonant and non-resonant with the A and B excitons, we observe a broad ground state bleach around 2.9 eV, with decay components similar to A and B. Associating this bleach with the band nesting region between K and Γ in the band structure indicates significant k-space delocalization and overlap among excitonic wave functions identified as A, B, C, and D. Comparison of time dynamics for all features in resonance and non-resonance excitation is consistent with this finding.*

PACS numbers:

*Corresponding author: stinaff@ohio.edu

## I. INTRODUCTION

Two-dimensional crystals of transition metal dichalcogenides (TMD) have attracted great attention due to their diverse, and potentially useful, electronic and optical properties. Many TMD monolayer materials are excellent candidates for ultrasensitive photo detectors[1], valleytronics[2], and energy harvesting devices[3]. For example, monolayer $MoS_2$ has been used to fabricate low power field-effect transistors[4], logic circuits[5], and phototransistors[6]. Such applications will be aided through a clear understanding of the excitonic band structure, multi excitonic states, and carrier dynamics. Studies have shown that the uv-vis absorption spectrum of the material has at least three peaks commonly labeled as A, B, and C[7, 8, 9]. The absorption spectrum is typically dominated by the C peak which has been attributed to a band nesting transition arising from a maxima in the joint density of states (JDOS) (Fig. 1)[10, 11, 12]. Interestingly, the feature at C is at higher energy than the predicted band gap and therefore, though the absorption is exciton like the state is more of a resonance within the continuum (Fig. 1). This peculiar feature clearly shows a strong absorption with a potentially large binding energy, which nevertheless, due to the degeneracy with the continuum states is found to spontaneously decay[13]. This combination of characteristics may prove especially advantageous for ultra-sensitive photodetectors, high speed opto-electronics, and energy harvesting devices. Here, we present systematic measurements on mono-, few-, and multi-layer $MoS_2$ utilizing room temperature broadband femtosecond transient absorption spectroscopy (FTAS) which provides insight into the dynamics of multiple, potentially interrelated, processes. These results provide new information into the dynamics of the resonance at C and the interplay between the primary states A, B, and C, which indicate significant coupling between the states in K space.

Monolayer MoS$_2$ has been shown to be a direct gap semiconductor with intense photoluminescence and a large spin-orbit coupling[14,15]. The indirect to direct band gap crossover is caused by strong out-of-plane quantum confinement in a 7 Å thick monolayer and results in two efficient excitonic transitions, A and B, derived from the conduction band and spin-orbit split valance band around the *K* (*K′*) point in the Brillouin zone (Fig. 1)[16-19]. Strong and long range Coulomb interaction and weak screening in such 2D material make the excitonic states accessible at room temperature, with many studies observing an additional state, labeled C, with strong, broad absorption [7,8,9]. This state has been attributed to a transition arising from a maxima in the JDOS, not associated with an extrema in the quasi-particle band structure but rather from parallel conduction and valance bands in a localized region between K and Γ (band nesting) (Fig. 1)[10,11,12]. Plotting the difference between conduction band and valance band (CB-VB) reveals the optical band structure where C, as well as another minima labeled D, appear (Fig. 1)[13]. Klots, et. Al., have shown that, with the inclusion of excitonic effects, minima such as C and D are expected to have bound states shifted by a binding energy on the order of 500 meV below the optical band structure minima, similar to what is observed for the A and B exciton peaks[13]. Using photoconductivity to measure the electric field required to ionize the various exciton states, this previous work revealed that absorption at C could result in an unbound electron-hole pair even in the absence of an external field, which was explained by the energetic degeneracy of C and the continuum[13]. This implies that despite the difference in k-space, C is delocalized enough to overlap with the continuum states near the $K$ ($K'$) point in the Brillouin zone resulting in efficient relaxation. This should also result in an overlap, and interaction, with the A and B excitons. In this work, we employ broadband FTAS to acquire time resolved measurements of the change in absorption in an energy range spanning A, B, C, and D resulting from an energy-tunable pump pulse. Selective excitation of the main absorption features gives unique insight into their interdependence and the underlying excitonic features of the material.

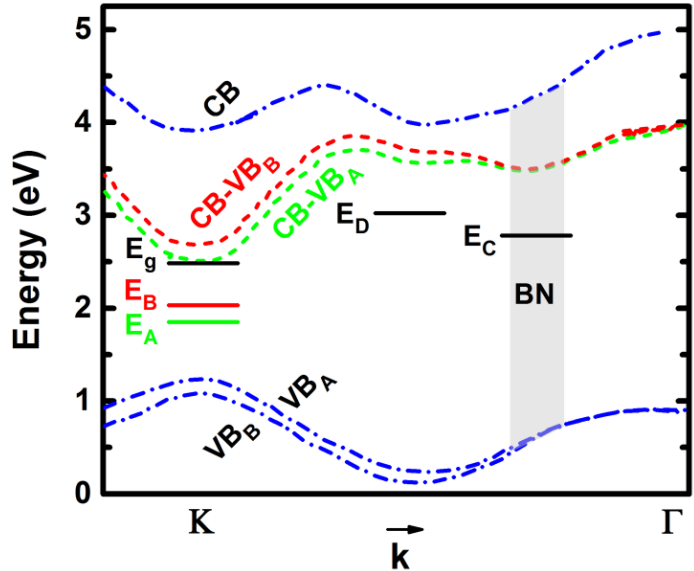

FIG. 1. (color online). Quasi-particle band structure of 1L MoS$_2$ (blue dotted lines), and the optical band structure (CB-VB, red and green dashed lines) created after A. R. Klots, et. Al. The optical band gap, $E_g$, is shown as well as the excitonic states, $E_A$, $E_B$, $E_C$, and $E_D$. It can be seen that $E_D$ is associated with a conduction band minimum in the quasi-particle band structure whereas, $E_C$ arises from band nesting (BN) due to the parallel nature of the conduction and valence band.

## II. SAMPLES AND EXPERIMENTAL METHODS

The primary samples used in this work were grown by Chemical Vapor Deposition (CVD) on sapphire and quartz substrates where $MoO_3$ and sulfur powder were placed inside a quartz tube inserted into a furnace heated under Argon flow at atmospheric pressure. A sapphire substrate, seeded with F19CuPC [20], was used to grow the first sample, at 650°C, which resulted in a thin film of $MoS_2$ with thicknesses varying between monolayer at the edges to 7 or more layers in the middle [Fig. 2(a) inset]. The sample was characterized using Raman spectroscopy and displayed the expected Raman modes, $A_{1g}$ and $E_{2g}$, with typical 20 $cm^{-1}$ separation for monolayer area as shown in [Fig. 2(a)]. The separation of the Raman modes increased as expected for the two layers, few layers and bulk areas. Absorption and PL spectra taken using a confocal microscope system (Witec) with 50 x objective lens and 0.85 numerical aperture at room temperature with 0.9 mW, 532 nm, excitation for PL and a Halogen white source for the absorption. PL showed the expected two excitonic peaks A (1.87 eV) and B (2.04 eV) [Fig. 2(b)]. Atomic Force Microscopy (AFM) was used to confirm the thickness of monolayer area to less than 1 nm [Fig. 2(b)]. The second sample has monolayer and few layers crystalline triangles grown on quartz substrate under 830°C for 10 minutes [Fig. 2(c)]. As before, the sample was characterized using Raman, Photoluminescence spectroscopy, and AFM [Fig. 2(C-d)], where we observe similar behavior to that of the first sample.

Ultrafast Transient absorption data have been collected using a femtosecond

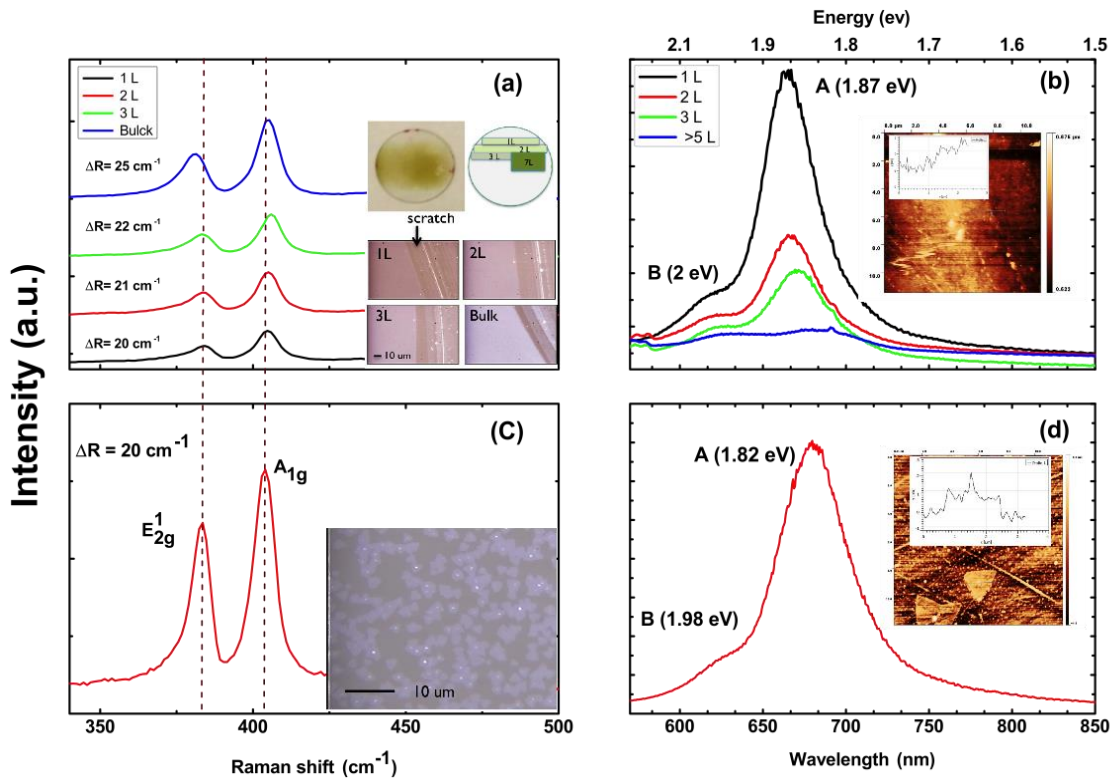

FIG. 2. (color online). (a) Raman signals of 1L, 2L, 3L, bulk $MoS_2$ on sapphire substrate. inset: An optical image for the CVD $MoS_2$ film on sapphire substrate with sketch of several positions with different numbers of layer, and Optical contrast for different positions on the sample with different numbers of layer. (b) PL emission for previous positions. Inset, AFM image for monolayer area with less than 1 nm thickness. (c) Raman signal for $MoS_2$ 1L triangle on quartz substrate. Inset: Optical image for the sample. (d) PL emission from monolayer triangle. Inset: AFM image of single layer triangle.

Ti:Sapphire one box amplifier (Spectra Physics Solstice 1 kHz, 4.1 W, 100 fs, 795 nm). The pump output passes through a 50/50 beamsplitter from which 50% is directed into an OPA (Light Conversion Topas-C) equipped with two Second Harmonic and two Sum Frequency crystals which can be tuned form 2200 nm to 250 nm while the other 50% passes through a second beamsplitter from which 95% is dumped and 5% is directed into a spectrometer for white light generation. The spectrometer (Ultrafast Systems Helios) used for the experiments excites the samples with a pump pulse generated from the OPA and probes the sample with a white light continuum generated by passing the 5% of the 795 nm pulse through a $CaF_2$ crystal (330 nm – 650 nm) or Sapphire crystal (430 nm - 750 nm). The white light is delayed relative to the pump pulsed using a motorized translation stage. Typical probe beam size is approximately 0.2 mm in diameter and is slightly smaller than the pump beam size. Typical experiment conditions use excitation pump power form from 50 µW to 700 µW.

Transient data is corrected and analyzed using Surface Xplorer Pro 1.1.5 software (ultrafast Systems) by subtracting spectral background features, applying nonlinear chirp correction, $t_0$ correction, single wavelength kinetic fitting and global fitting from single value decomposition procedure. Additional details of the unmodified instrument are reported elsewhere[21]. The data for the FTAS are reported as a differential absorption, $\Delta A = A(t) – A(0)$, where $A(0)$ and $A(t)$ are the absorption spectra measured before, and a time $t$ after, the pump pulse, respectively. In general, features in the differential absorption spectra may arise from a combination of changes in the intensity, linewidth, or position of features in the unperturbed absorption spectrum. For example, broadening of the absorption spectra, while maintaining the total area of the absorption peak, would give rise to both positive and negative $\Delta A$ features[22]. A positive $\Delta A$ may also be attributed to photo-induced absorption (PIA), indicating the availability of new absorption pathways in the excited system, and therefore an increased absorption relative to the ground state or unexcited $MoS_2$. A negative ΔA may indicate either a reduction of the available ground state carriers due to excitation from the pump, typically referred to as a bleach, or stimulated emission of the pump-induced excited states.

### III.    RESULTS AND DISCUSSION

Figure 3 shows a comparison of the absorption, PL, and transient absorption spectrum obtained at a pump-probe delay of 1 ps. The transient absorption spectrum of monolayer film $MoS_2$ [Fig. 3 (a)] shows at least seven major features, three positive ΔA features, and four bleaches. The positive ΔA features between 2.1 eV and 2.7 eV result from excitation dependent broadening of the main bleach features[22]. This is consistent with pump-power dependent FTAS where we observe a shift of the features between 2.1 eV and 2.7 eV while all other maxima remain fixed (Fig. 4). A simple modeling of the ΔA spectra demonstrates that a modest broadening of less than a percent gives rise to positive ΔA features consistent with the measured spectra. This broadening effect may explain the positive ΔA peaks around (1.96 eV, 634 nm), and (1.79 eV, 692 nm) [Fig. 3 (a)], however, these may also be a result of AB and AA biexcitons respectively, consistent with previous studies[23].

Comparing the absorption and transient absorption spectra of monolayer $MoS_2$ [Fig. 2 (c)], we find that the three bleach signals align well with the observed absorption peaks. Both spectra are dominated by a high-energy feature around 2.9 eV[24,25], as well as two smaller features at 1.88 eV and 2.04 eV, which correspond to the A and B excitons. The A and B peaks are the well-studied excitonic states derived from the spin-orbit split

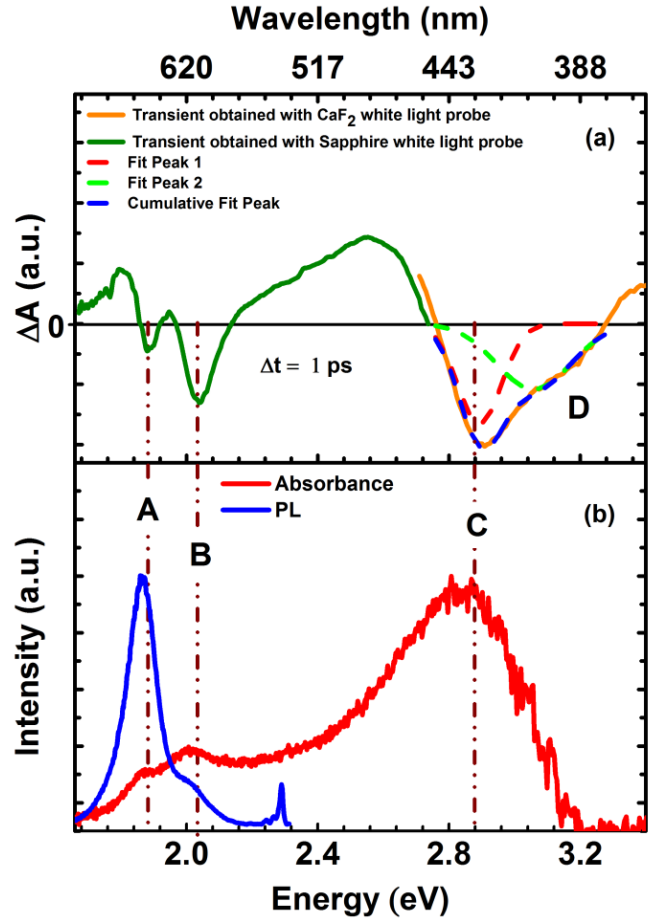

FIG. 3. (color online). (a) Transient absorption spectrum of monolayer $MoS_2$ taken at a pump-probe delay of 1 ps with 3.1 eV excitation. Fits the broad bleach C. (b) Absorption and Photoluminescence spectra for the same monolayer $MoS_2$.

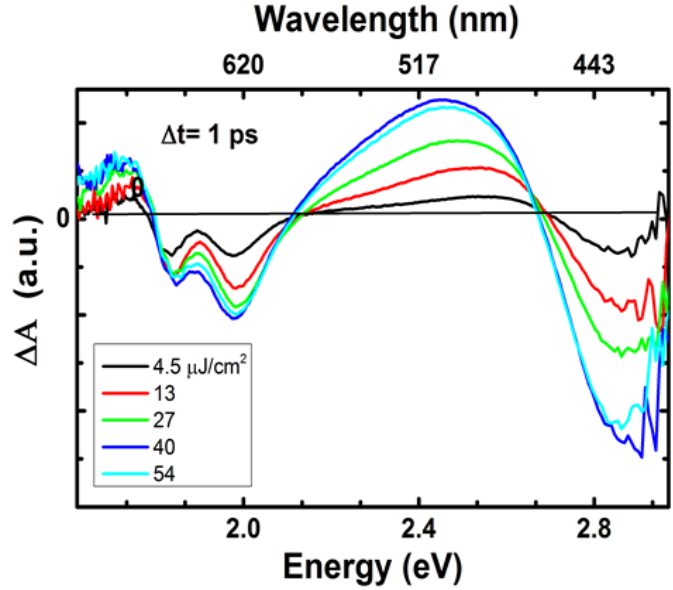

FIG. 4. (color online).   Power dependence of transient absorption spectra under nonresonant excitation

valance band and the degenerate conduction band[14,15]. When using a blue shifted probe (CaF$_2$ generated white light continuum) we also observe that the feature around 2.9 eV is clearly composed of two bleaches, labeled C and D. As described earlier, the feature at C is due to locally parallel bands between K and Γ giving rise to a local minima in the optical band structure as well as a singularity in the JDOS. The bleach at D is likely an excitonic state associated with either the higher energy local minimum in the conduction band near C, or an additional band nesting, resulting in a minimum in the optical band structure (Fig. 1)[10,13]. ΔA spectra were collected at three different pump energies, one high energy and non-resonant with any observed feature (3.1 eV, 405 nm) and two energies resonant with the B (2.04 eV, 608 nm), and A (1.88 eV, 660 nm) exciton states. For the two longer wavelength excitations we increased the energy density of the pump by about one order of magnitude to compensate for the decrease in absorption[8]. Surprisingly, the ΔA spectra for the three excitation energies yield similar results in terms

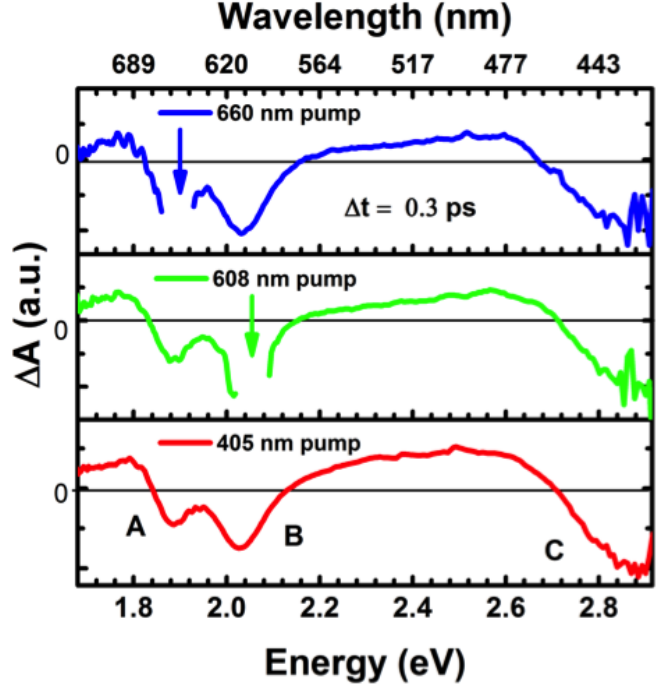

FIG. 5. (color online). Transient absorption spectra for monolayer MoS$_2$ all taken at a pump-probe delay of 0.3 ps with three different excitation energies, one non-resonant (3.1 eV, 405 nm), one in resonance with B, and one in resonance with A. We note that the prominent bleach features are present for each excitation energy

of the presence of the A, B, and C bleaches (Fig. 5). The bleach in B while pumping in resonance to A has been previously explained as an extremely fast spin relaxation between $K$ and $K'$ valleys and inter-valley coupling of the electronic part of the exciton wave function due to the degeneracy of the spin states. However, the unexpected presence of the higher energy C bleach when pumping in resonance with the low energy A or B states implies that the three absorption transitions are coupled. Considering the significant separation in k-space between C and the $K$ ($K'$) point, around which A and B are centered, a decrease in absorption at C, while resonantly creating A or B, indicates that the wave functions must consist of states with momenta spanning $K$ ($K'$) and $\Gamma$ (Fig. 6). This is consistent with the highly localized real space wavefunction describing the A and B excitons[26,27], the predicted localized nature of the C resonance, as well as the spontaneous decay of the C state into free electron-hole pairs, indicating an overlap of the C wavefunction with the continuum in k-space.

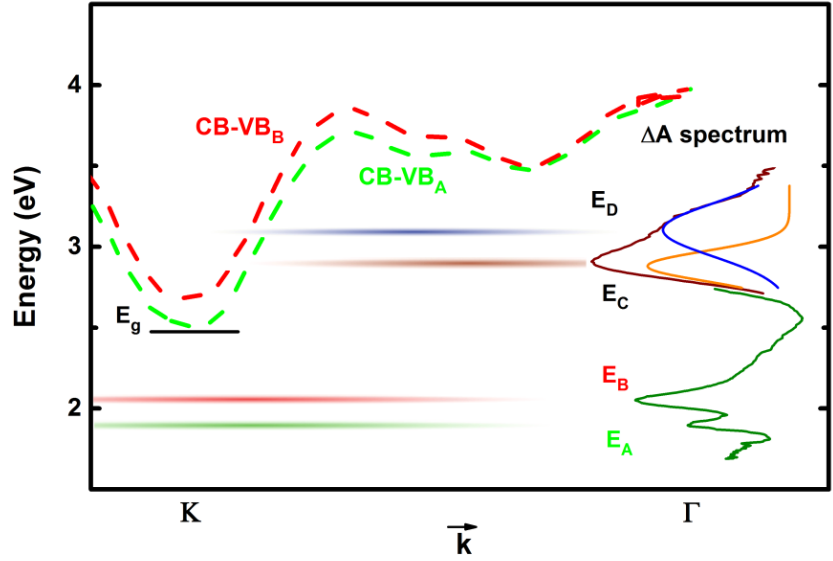

FIG. 6. (color online). The optical band structure reproduced after A. R. Klots, et. Al and the ΔA spectrum of 1L $MoS_2$, A, B, and C wavefunctions must consist of states with momenta spanning K (K′) and Γ, the second peak within C bleach indicates forth excitonic state D.

Comparison of ΔA and absorption spectra for mono-, few-, and multi-layer $MoS_2$ regions (Fig. 7) show that A, B, C, and D red shift by different amounts as the number of layers increases. This red shift supports the conclusion that the bleaches are derived from different regions in the Brillouin zone which are affected to different degrees by the reduced confinement, and inter-layer coupling as layer number increases[28,29,30]. Specifically, C has the largest red shifted with increasing layer number while A and B have the smallest, in agreement with the theoretical prediction for the transitions around Γ to be more affected by changes in confinement and inter layer-coupling.

The transient absorption kinetics (Fig. 8) was fitted to a fourth order multiexponential decay at single wavelengths and by singular value decomposition (SVD). The results for 3.10 eV excitation are listed in Table 1, the results for 1.84 eV excitation are listed in Table 2. Interestingly, the time constants are in agreement with one another between the two excitation wavelengths where we see, for A, B, and C, time constants in the range of 0.7 ps, 5 ps, 30 ps, and 240 ps at 3.1 eV excitation. This agreement supports our picture that the resonant and non-resonant excitations perturb the entire ground state including all excitonic transitions and therefore the measured time constants represent the recovery of the global ground state after the excitation. This also indicates that any effect of the continuum carriers generated by resonant absorption into the C peak is faster than the temporal resolution of our experiments. The typically low quantum yield of the material implies that nonradiative mechanisms dominate the decay dynamics[14]. Comparison between multiple previous studies reveal a range of reported exciton dynamics with multi exponential decays and time constants ranging from a few hundred femtoseconds to a few nanoseconds[31, 32, 33, 34, 35, 36]. For two-dimensional materials, where defects and surface interactions may dominate, such variance arising from sample and experimental conditions is perhaps not surprising.

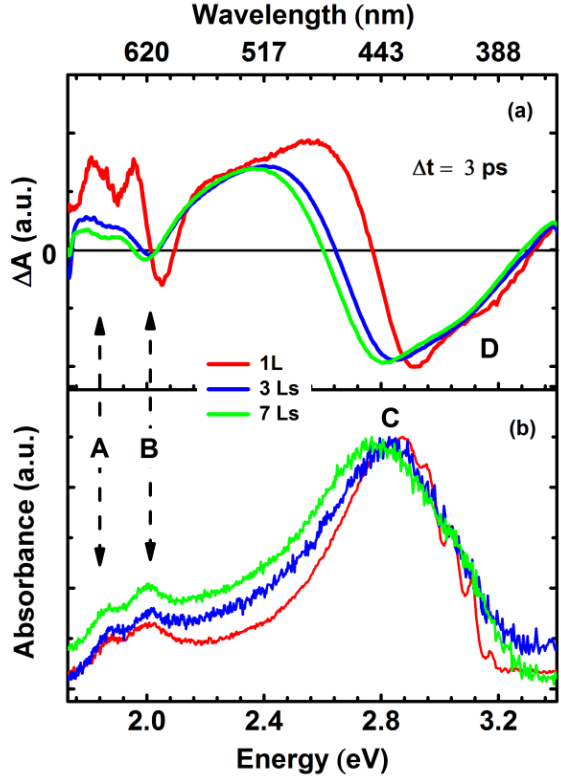

FIG. 7. (color online). (a) Transient absorption spectra for varying number of layers (1, 3, and 7) of $MoS_2$ taken at a pump-probe delay of 3 ps. A small red shift for each peak position with respect to the monolayer spectrum is observed. (b) Absorption spectra for the same layer numbers as in (a) showing a similar shift.

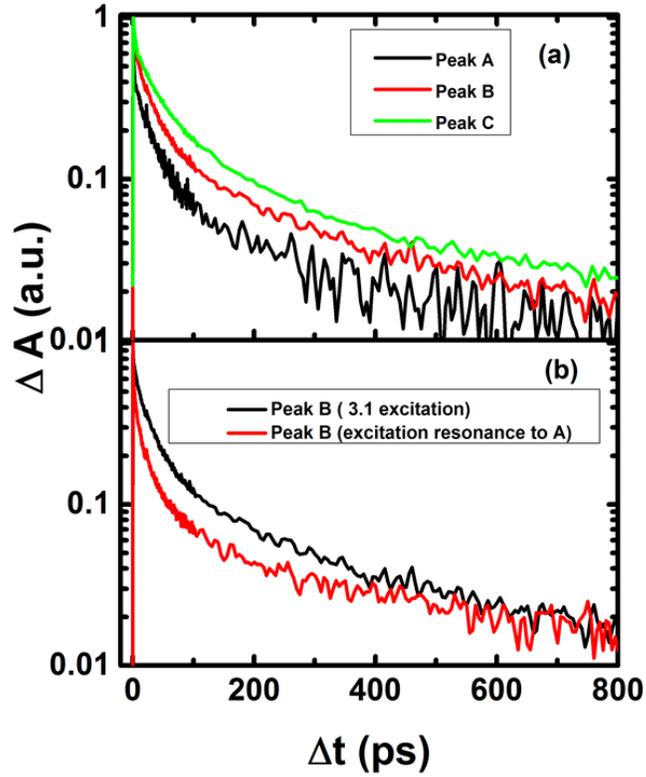

FIG. 8. (color online). (a) Decay curves for the three bleaches A, B, and C showing similar dynamics. (b) Decay curves for B bleach with 3.1 ev excitation and excitation resonance to A , these data are related to $MoS_2$ tringles on quartz substrate.

The sub-ps decay of the bleach features correlates well with the sub-ps rise time of the positive ΔA features between B and C, and is consistent with our interpretation of these positive features as arising from broadening. The mechanism responsible for this component is present for both resonant and non-resonant excitation, indicating that it is not likely associated with free carrier interactions and may be related to defects or other intrinsic properties in our CVD grown material. Bi-exciton formation could also contribute to this component. At room temperature, the fast $t_1$ component has typically been assigned to Auger scattering [31,35]. The variance in this component in the literature could be related to experimental time resolution, and indicates the extent to which defects play a role in the time dynamics between different samples. The second component may be assigned to coupling to surface trap states leading to the rapid relaxation of the excited carriers as well as exciton–exciton interactions, as shown by Sun et al., which at high excitation densities in the range of $10^{12}$ cm$^{-2}$ may give rise to a few ps decay component. Non-radiative decay through carrier-phonon and slow Auger scattering are consistent with the $t_3$ component[31,33]. The final component is related to radiative recombination, which has been reported to be hundreds of picoseconds, to nanoseconds, at room temperature[33, 34, 36]. Observation of the final time components reveals that they account for less than 10% of the signal intensity on average which matches well with the low quantum yield of emission often observed from these monolayer materials.

Table 1 Fitting results for the FTAS experiment with 400 nm excitation. Note that the rise of the broadening feature at 475 nm is comparable to the initial decay of C and the exciton bleaches at A and B.

| Peak \ t (ps) | C | B | A | AA | AB | Broadening (475 nm) | Global fit |
|---|---|---|---|---|---|---|---|
| $t_1$ | 0.72 ± 0.21 (9.9%) decay | 0.70 ± 0.0 (2.8%) decay | 0.70 ± 0.00 (31.6%) decay | 1.00 ± 0.05 (36.1%) rise | 0.53 ± 0.04 (30.6%) rise | 1.04 ± 0.03 (44.9%) rise | 1.05 ± 0.09 |
| $t_2$ | 5.72 ± 0.60 (41.4%) decay | 6.64 ± 0.79 (41.4%) decay | 4.36 ± 0.43 (43.4%) decay | 8.33 ± 2.69 (12.9%) decay | 2.01 ± 0.11 (33.9%) rise | 12.49 ± 8.3 (8.4%) decay | 6.52 ± 1.37 |
| $t_3$ | 31.17 ± 3.24 (39.2%) decay | 33.51 ± 3.56 (48.6%) decay | 23.59 ± 3.39 (22.2%) decay | 35.39 ± 3.8 (38.9%) decay | 41.29 ± 1.73 (26.7%) decay | 37.4 ± 6.97 (35%) decay | 32.9 ± 5.8 |
| $t_4$ | 242 ± 44 (9.5%) decay | 250 ± 42 (12.8%) decay | 242 ± 99 (2.8%) decay | 246 ± 30 (12.2%) decay | 270 ± 26 (8.9%) decay | 251 ± 37 (11.7%) decay | 231 ± 39 |

Table 2 Fitting results for the FTAS experiment at excitation resonance to A. Note that the fitting at A is not reliable since it is coincident with the excitation.

| Peak<br>t (ps) | C | B | A | AA | AB | Broadening<br>(475 nm) | Global fit |
|---|---|---|---|---|---|---|---|
| $t_1$ | 0.41 ± 0.21<br>(14.1%) decay | 0.46 ± 0.12<br>(22.2%) decay | X | 0.88 ± 0.14<br>(37.2%) rise | 0.93 ± 0.05<br>(71.5%) rise | 0.84 ± 0.09<br>(43.5%) rise | 1.05 ± 0.09 |
| $t_2$ | 4.2 ± 0.59<br>(42.8%) decay | 4.01 ± 0.59<br>(41.4%) decay | X | 6.39 ± 1.67<br>(17.9%) decay | 5.0 ± 0.00<br>(1.2%) rise | 5.0 ± 00.0<br>(7.7%) decay | 6.52 ± 1.37 |
| $t_3$ | 30.47 ± 5.39<br>(33.4%) decay | 30.13 ± 5.46<br>(28.4%) decay | X | 29.61 ± 6.78<br>(33.6%) decay | 29.91 ± 10.22<br>(19.4%) decay | 26.58 ± 6.85<br>(35.3%) decay | 32.9 ± 5.8 |
| $t_4$ | 342 ± 115<br>(9.7%) decay | 361 ± 136<br>(8%) decay | X | 254 ± 73<br>(11.6%) decay | 207 ± 75<br>(7.9%) decay | 216 ± 56<br>(13.6%) decay | 231 ± 39 |

## IV. CONCLUSION

Using broadband FTAS on monolayer $MoS_2$, with selective excitation, we have observed a strong correlation between the well-studied A and B excitonic peaks and the high-energy absorption feature associated with band nesting transitions. We have observed that, even when resonantly exciting the low energy exciton peaks, the feature in the ΔA spectra at C is still present. This implies a significant delocalization in k-space resulting in substantial overlap between the A, B, C and D states. Despite the resonant nature of the C feature with the continuum, the temporal behavior is observed to be similar between each of these features. Such strong interdependence between the excitonic features in $MoS_2$ may provide a route for ultra-fast transmission modulation based on non-resonant excitation.

Acknowledgments. JJR thanks the National Science Foundation under Grants CHE 0809699, 0947031, and 1112250 for financial support of this work.